\documentclass[twocolumn,preprintnumbers,amssymb,aps,pra]{revtex4-1}
\usepackage{graphicx}
\usepackage{dcolumn}
\usepackage{bm}
\usepackage{amssymb}
\usepackage{amsmath}
\usepackage{supertabular}
\usepackage{longtable}
\usepackage{multirow}\usepackage{braket}
\usepackage{esvect}
\usepackage{gensymb}
\usepackage{soul}
\let\oldAA\AA
\renewcommand{\AA}{\text{\normalfont\oldAA}}

\usepackage{qtree}
\usepackage{tipa}
\usepackage{appendix}
\usepackage{ulem}
\usepackage{xcolor}
\usepackage{tabularx,booktabs}

\usepackage[colorinlistoftodos]{todonotes}
\presetkeys{todonotes}{inline,backgroundcolor=white}{}

\begin{document}

\title{Digital quantum simulation of strong correlation effects with iterative quantum phase estimation over the variational quantum eigensolver algorithm: $\mathrm{H_4}$ on a circle as a case study}
\author{Dipanjali Halder$^{1,*}$, V. S. Prasannaa$^{2,*}$, Valay Agarawal$^1$, Rahul Maitra$^{1,\dagger}$}
\affiliation{$^1$ Department of Chemistry, \\ Indian Institute of Technology Bombay, Powai, Mumbai 400076, India \\
$^2$ Centre for Quantum Engineering, Research and Education, TCG CREST, Salt Lake, Kolkata 700091, India\\
$^*$ These authors have contributed equally to this work\\
$^\dagger$ rmaitra@chem.iitb.ac.in}

\date{\today}

\begin {abstract}
The iterative quantum phase estimation algorithm, applied to calculating the ground state energies of quantum chemical systems, is theoretically appealing in its wide scope of being able to handle both weakly and strongly correlated regimes. However, the \textit{goodness} of the initial state that is sent as an input to the algorithm could strongly decide the quality of the results obtained. In this work, we generate the initial state by using the classical-quantum hybrid variational quantum eigensolver algorithm with unitary coupled cluster ans\"{a}tz. We apply the procedure to obtain the ground state energies of the $H_4$ molecule on a circle, as the system exhibits an interplay of dynamic as well as static correlation effects at different geometries. Furthermore, we argue on the importance of static correlation in construction of the reference determinant, and propose a minimally parametrized unitary coupled cluster ans\"{a}tz, which drastically reduces number of variational parameters while incorporating the static correlation effects in the wavefunction. We demonstrate that a carefully and appropriately prepared initial state can greatly reduce the effects of noise due to sampling in the estimation of the desired eigenphase.  
\end{abstract}

\maketitle

\section{Introduction}\label{sec1} 

Quantum computers offer great promise in efficiently handling certain problems that are otherwise intractable on classical devices~\cite{shor,dj,hhl}. Over the last two decades, the emerging field of quantum science and technologies has witnessed remarkable progress with applications spread across physics, chemistry, mathematics,  biology, and finance~\cite{math,bio,fin}. In particular, novel approaches to calculate ground state energies of quantum chemical systems, and which promise an exponential speed-up over classical devices, have made these class of problems a killer application of quantum computing (for example, see Ref.~\cite{ag,ag1,ag2,Whitfield}). \\

In a seminal work, Aspuru Guzik \textit{et. al.} carried out a digital quantum simulation of two molecular systems, using the full configuration interaction (FCI), and computed their ground state energies~\cite{ag}. The approach was based on the famous quantum phase estimation (QPE) algorithm of Abrams and Lloyd~\cite{Abrams and Lloyd 1997,Abrams and Llyod 1999}. The QPE algorithm will lead to a significant reduction in time complexity as compared to classical devices in the future when many more qubits are available. \\

Based on a complementary paradigm, the variational quantum eigensolver (VQE) algorithm is a hybrid quantum-classical algorithm, which 
aims to find the lowest
eigenstate of the interacting many-body Hamiltonian by varying the energy expectation value function with respect to the parameters in the wavefunction ans\"{a}tz~\cite{Peruzzo 2014,McClean 2016}. The method relies upon the Rayleigh-Ritz variational principle~\cite{Griffiths} and therefore provides an upper bound to the exact eigenenergy of the many-body Hamiltonian. It is worth adding at this point that the choice of a good ans\"{a}tz can, in general, give results with excellent precision. One such widely used variational form is the chemically motivated unitary coupled cluster (UCC) ans\"{a}tz. The coupled cluster approach, which is widely regarded as the gold standard of electronic structure calculations, is known to efficiently capture dynamical correlation in a system. However, in the regions of molecular strong correlation, which is attributed to the quasi-degeneracy of multiple constituent determinants, the traditional VQE algorithm may not fare well, and as a result, suffer from  inconsistent description over the molecular potential energy surface (PES)~\cite{Sugi1}. On the other hand, the QPE algorithm does not possess the limitation if appropriate care taken to prepare the initial state, as we shall discuss next. \\

The QPE algorithm~\cite{qpe_terhal} can be thought of as a projection of an input state (called the reference state or the initial state interchangeably) onto an eigenstate, whose eigenvalue we then compute with the algorithm. This necessitates the initially prepared reference state having at least a non-negligible overlap with the exact eigenstate of the Hamiltonian, and hence
preparation of a \textit{good} initial state is of utmost importance. There have been a number of works in literature that have addressed the issue of constructing an initial state that has substantial overlap with the eigenstate of the system  Hamiltonian. In the molecular weak correlation regime, the simplest choice, namely the single determinant Hartree-Fock (HF) function, often serves this purpose quite well. However, this is often not the case while dealing with strongly correlated systems, and in such scenarios, using a combination of multiple Stater determinants as the initial 
reference function becomes very important. This line of thought has led various authors to construct the initial state by various means, for example, by employing the classically computed Complete Active Space Self-Consistent Field (CASSCF) method~\cite{Pittner, Hoffmann} or using Adaptive Sampling CI~\cite{refhttps://arxiv.org/abs/1809.05523} techniques. Both of these approaches have shown substantial squared overlap of the initial reference state with the eigenstate of the Hamiltonian, and thus a high probability of success for the QPE algorithm. While the CASSCF method is extremely powerful and useful for strongly correlated molecular systems, the associated computational cost increases steeply against the size of the active space. Thus, it is quite important to develop a wavefunction preparation technique that should lead to handling the strong and weak correlation effects in a balanced manner, and with relatively low computational scaling. Sugisaki \textit{et. al.}~\cite{(doi/10.1021/acscentsci.8b00788)} have computed the reference state by exploiting the diradical characters of the molecules from spin-projected Unrestricted HF  wavefunctions. This method entirely bypasses any post-HF calculations for the 
reference state preparation, and drastically improves the success probability of QPE, particularly for molecules with intermediate diradical characters. A complementary 
approach for initial state preparation is often followed, where the HF Hamiltonian is slowly and systematically changed over to the FCI Hamiltonian, such that the system remains in its instantaneous ground state. This is known as the Adiabatic State Preparation (ASP) in  literature~\cite{ag,Sugi2,Sugi3}. \\ 

In this work, we employ an iterative version of QPE algorithm (IQPE)~\cite{iqpe,iqpe1}, and opt for an alternative way for preparing its input reference state by using the VQE algorithm with the UCC ans\"{a}tz. We refer to this approach hereafter as IQPE over VQE. A schematic of the approach is given in Fig.~\ref{one}. We pick IQPE over traditional QPE for our simulations, in view of many ancilla qubits that are required for the latter, which in turn makes computations very expensive. To the best of our knowledge, there is one other work in literature, where the authors independently suggest QPE over VQE, but they do not study the effect of various VQE parameters in the initial state preparation~\cite{Romero}. In this work, we systematically tune the number of VQE iterations, and study its connection to the  probability of successfully landing on the right state when fed as an input to the IQPE algorithm. We will demonstrate that a multi-determinantal reference state meticulously prepared using the VQE algorithm is likely to have a superposition of all the dominant determinants of the FCI wavefunction, and thus we anticipate that one can systematically tune it to have considerable overlap with the exact eigenstate of the Hamiltonian, even in strong correlation regimes. We then demonstrate the strength and applicability of this method by carefully selecting a physical system that displays varied electronic complexity across its PES, that is, the system should be dominated by strong correlation effects in some geometries and weak in others to serve as a good testbed for the algorithm. After generating the PES and studying the precision of our final results with the number of IQPE iterations, we show that our approach provides a good and systematically improvable probability of success to land on the correct state. We would also introduce a minimally parametrized CC ans\"{a}tz, where the multi-determinantal reference function is generated by incorporating the static correlation effects via a unitary ans\"{a}tz with \textit{internal} excitations. This approach promises to be a good compromise between the accuracy in terms of the squared overlap and the number of variational parameters, while constructing the reference state. We finally extend our analyses by studying the scenario with Qiskit's QASM backend, where sampling comes into play. \\

The paper is organised as follows: we discuss in Section \ref{theory} the VQE, QPE, and the IQPE algorithms, followed by the hybrid IQPE over VQE approach. This is followed by the details of our computations and the rationale behind our choice of system. Finally, we present our obtained results and the analyses that follow from our data in Section \ref{results}, where we systematically expound details such as the PES, effect of IQPE iterations in precision of our results, the effect of VQE iterations on state preparation for IQPE, he minimally parametrized ans\"{a}tz, and finally analysis of our results by taking into account sampling. We conclude in Section \ref{conclusion}. \\

\section{Theory}\label{theory}
\subsection{The Variational Quantum Eigensolver algorithm} 

As mentioned earlier, the classical-quantum hybrid VQE algorithm relies upon the variational principle , and hence the ground state energy, $E(\theta)$, given by 

\begin{eqnarray}
E(\theta) &=& \frac{\langle \Psi(\theta)|H|\Psi(\theta) \rangle}{\langle \Psi(\theta)|\Psi(\theta) \rangle} \\
&=& \langle \Psi_{HF} | U(\theta)^\dagger H U(\theta) | \Psi_{HF} \rangle, \label{eqE}
\end{eqnarray}

is, in principle, guaranteed to be an upper bound to the true ground state energy. Here, the state, $|\Psi(\theta) \rangle$, is expressed in terms of a unitary operator, $U(\theta)$, which is parametrized by a set of parameters concisely denoted as $\theta$, and $U(\theta)$ acts upon a reference state, $|\Psi_{HF} \rangle$, which is the HF state in this work. We choose $U(\theta)$ to be the UCC ans\"{a}tz with double excitations (UCCD) unless mentioned otherwise, and it is given by 

\begin{eqnarray}
\ket{\Psi(\theta)}&=&e^{{{T_2}({\theta})}-{T_2}^{\dagger}({\theta})}\ket{\psi_{HF}}; \label{excop} \\
{T_{2}}({\theta})&=&\frac{1}{2}\sum \limits_{i,j;a,b}\theta_{ij}^{ab}a_{b}^{\dagger}a_{a}^{\dagger}a_{j}a_{i}. 
\end{eqnarray}

Here, the subscripts $i,j$ refer to the occupied spin-orbitals in the HF reference determinant, while $a,b$ are the
unoccupied/virtual spin-orbitals. The t-amplitudes, $\theta_{ij}^{ab}$s, play the role of the VQE parameters. The choice to not consider single excitations has to do with the system that we work with, where we expect the double excitations to be dominant, and we shall explain and demonstrate this point in Section~\ref{results}. One may also choose only a selection of double excitations involving the chemically \textit{active} orbitals. This important point will be elaborated towards the end of the manuscript, in the context of the minimally parametrized ans\"{a}tz. Note that in Eq. (\ref{eqE}), the many-body molecular electronic Hamiltonian, $H$, can also be expressed in second quantized form as 

\begin{eqnarray}
H&=&\sum\limits_{p,q}h_{pq}a_{p}^{\dagger}a_{q}+\frac{1}{2}\sum\limits_{p,q,r,s}h_{pqrs}a_{p}^{\dagger}a_{q}^{\dagger}a_{r}a_{s} \\
&=& \sum_\alpha h_\alpha P_\alpha, \label{eqH}
\end{eqnarray}

where $h_{pq}$ and $h_{pqrs}$ refer to one- and two- electron integrals, respectively. 
The integrals are supplied as inputs to the VQE algorithm, using a program run on a 
classical computer. In Eq. (\ref{eqH}), the Hamiltonian is compactly described using a single index, $\alpha$, and the $P_{\alpha}$s no longer refer to the second 
quantized creation and annihilation operators, but their mapped version into their qubit 
operator form, that is, a string of tensor product of Pauli operators. This can be done by, for example, the Jordan-Wigner transformation~\cite{map}. $T_2(\theta)$ can also be appropriately mapped into its qubit operator form, and this allows $|\Psi(\theta) \rangle$ to be recast in its circuit form~\cite{Whitfield}, with appropriately chosen initial guess parameters, $\theta_{init}$. The expectation value is calculated by either using matrix 
operations (statevector (SV) backend) or the circuit is evaluated by measuring $H$ on
$|\Psi(\theta_{init}) \rangle$ (QASM backend) and the corresponding energy extracted. This sub-part constitutes the quantum module of the classical-quantum VQE approach. The energy is then minimized with respect to the parameters by using a suitable optimization algorithm on a classical computer, over a series of iterations, where in each VQE iteration, the parameter $\theta$ incrementally changes, and is fed back into the circuit to obtain an updated value of energy for that iteration, and the process is repeated until a minimum is found. \\

The excitation operators that occur in the exponent of Eq. (\ref{excop}), when acting on the reference HF determinant, capture the physical effects arising from the resulting determinants that differ from their HF counterpart by double excitations. With a sufficient number of VQE iterations for energy minimization, the HF reference function thus evolves to a correlated ground state wavefunction, which is expected to have a substantial overlap with the \textit{exact} ground eigenstate. While this expectation is indeed the case for molecules in weakly correlated regimes, the same is not strictly true for strong correlation, which we shall explain later with a concrete example, $H_4$ on a circle. This leads VQE to provide a poor estimation of the energy in the presence of molecular strong correlation, as we will demonstrate later. However, such a correlated wavefunction resulting as a byproduct of the iterative optimization of the energy may be used for QPE with a high probability of its success, even in in the regions of molecular strong correlation. \\

In the following subsection, we will briefly discuss the general principles of the QPE algorithm starting from an arbitrary reference function, and highlight the importance of reference state preparation in this regard. We will then demonstrate how the evolved state resulting from VQE energy optimization can suitably be fed as the reference state into the QPE algorithm. \\

\subsection{Quantum phase estimation algorithm}

The algorithm aims at estimating the energy eigenvalues of a physical Hamiltonian. The objective is to estimate the phase, $\phi$, of an eigenvector of a unitary operator, ${U}$, such that 

\begin{eqnarray}
{U\ket{\Psi}=e^{2\pi{i}\phi}\ket{\Psi}}.
\end{eqnarray}

The algorithm uses two registers. One of them  contains a set of $k$ ancilla qubits, prepared  initially in a state, $\otimes_{i=1}^k \ket{0}$, while the other one contains a set of qubits required to define the initial state of the system. This arrangement enables phase kickback via a series of controlled unitary gates. The inverse quantum Fourier transform is then employed to change bases. This step is  followed by extracting the phase, $\phi$. The precision of the algorithm depends on $k$.  However, in practice, the number of ancilla qubits which can be implemented is very limited, in view of the current Noise Intermediate Scale Quantum (NISQ) era that we live in, thus restricting the applicability of the promising approach to small systems and/or limiting the precision. An iterative version of QPE, namely the IQPE algorithm,  uses a single ancilla qubit to solve the eigenvalue problem. \\

\subsection{The iterative quantum phase estimation algorithm}

The IQPE algorithm carries out the conventional QPE with only a single ancilla qubit. The precision of the obtained results no longer depends on the number of ancilla qubits, but on the number of iterations. IQPE too requires two registers. However, the first register contains only a single ancilla qubit, initially in the state $\ket{0}$, and the second register contains the qubits necessary for defining the initial state of the system, $\ket{\Psi_{init}}$. \\

The phase, $\phi$, which is to be estimated, can be written down as a series of bits where each bit, $\phi_{i}\epsilon\{0,1\}$. 

\begin{eqnarray} \phi=\frac{\phi_{1}}{2}+\frac{\phi_{2}}{4}+\hdots+\frac{\phi_{m}}{2^{m}}=0.\phi_{1}\phi_{2}\hdots\phi_{m}. \end{eqnarray}

The algorithm starts by estimating the least significant bit, $\phi_{m}$, for which a Hadamard gate is applied to the ancilla qubit, $\ket{q_{0}}$, so that it is transformed to $\ket{+}$. This is followed by a controlled-$U^{2^{m-1}}$ operation between the ancilla qubit register ($q_0$) and the state register ($q_1$), such that the combined state of the ancilla and the state register becomes 

\begin{eqnarray}
\ket{q_{0}}\ket{\Psi_{init}}\longrightarrow
 \ket{0}\ket{\Psi_{init}}+e^{i2\pi 0.\phi_{m}}\ket{1}\ket{\Psi_{init}}.  
\end{eqnarray}

Hence, the phase that is kicked back into the ancilla is $0.\phi_{m}$. As $\phi_{m}$ is either 0 or 1 , a measurement on ancilla will lead to $\ket{+}$ if $\phi_{m}=0$, and $\ket{-}$ if $\phi_{m}=1$. This would produce outcomes $\ket{0}$ and $\ket{1}$, respectively, when measured in the x-basis. After the measurement of the least significant bit, $\phi_{m-1}$ is measured, using the value of $\phi_{m}$. For estimation of $\phi_{m-1}$, $2^{m-2}$ controlled-U operations are applied between $q_{0}$ and $q_{1}$ such that 

\begin{eqnarray}
    \ket{q_{0}}\ket{\Psi_{init}}\longrightarrow \ket{0}\ket{\Psi_{init}}+e^{i2\pi 0.\phi_{m-1}}e^{i2\pi \frac{\phi_{m}}{4}}\ket{1}\ket{\Psi_{init}}. \nonumber \\
\end{eqnarray}

Thus, the phase that is kicked back is $0.\phi_{m-1}+\frac{\phi_{m}}{4}$. To extract the bit $\phi_{m-1}$, a phase correction of $(-2\pi\phi_{m}/{4})$ is applied with the help of rotation gates, which transform the ancilla into $\ket{0}+e^{i2\pi 0.\phi_{m-1}}\ket{1}$. The ancilla is then measured in the x-basis to extract $\phi_{m-1}$. Similarly, the more significant phase bits are measured using the knowledge of the previously measured lesser significant phase bits, and the process is iterated backwards till all the phase bits are recovered. \\

The above algorithm works even if the initial state $\ket{\Psi_{init}}$ on the $q_{1}$ register is not an eigenfunction of the unitary operator, U but a linear superposition of various eigenfunctions ~\cite{Cruz},

\begin{eqnarray}
    \ket{\Psi_{init}}=\sum\limits_{n}c_{n}\ket{u_{n}}.
\end{eqnarray}

In such a situation, the algorithm will lead us to one of the eigenfunctions, with a probability $|{c_{n}}|^{2}$ and provide an estimate of the local phase of the eigenfunction $\ket{u_{n}}.$ 
In our present work, we would explore two different unitary ans\"{a}tze to construct multi-determinantal reference state via VQE, and we will show that a careful construction of the reference state is essential for the success of the IQPE algorithm. \\

An important point to note is that the initial state does not influence the precision of the phase, rather it only affects the probability with which the phase of a particular eigenstate is measured. The precision of the algorithm itself is solely determined by the number of iterations in IQPE. \\

\section{Results and Discussions}\label{results}

\begin{figure*}[t]
    \centering
    \setlength{\tabcolsep}{1mm}
        \begin{tabular}{cc}
            \includegraphics[height=40mm,width=40mm]{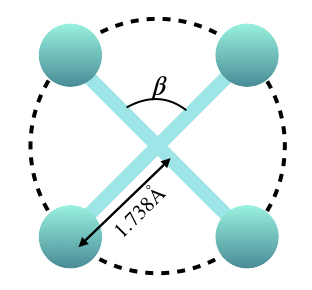} &
            \includegraphics[height=60mm,width=120mm]{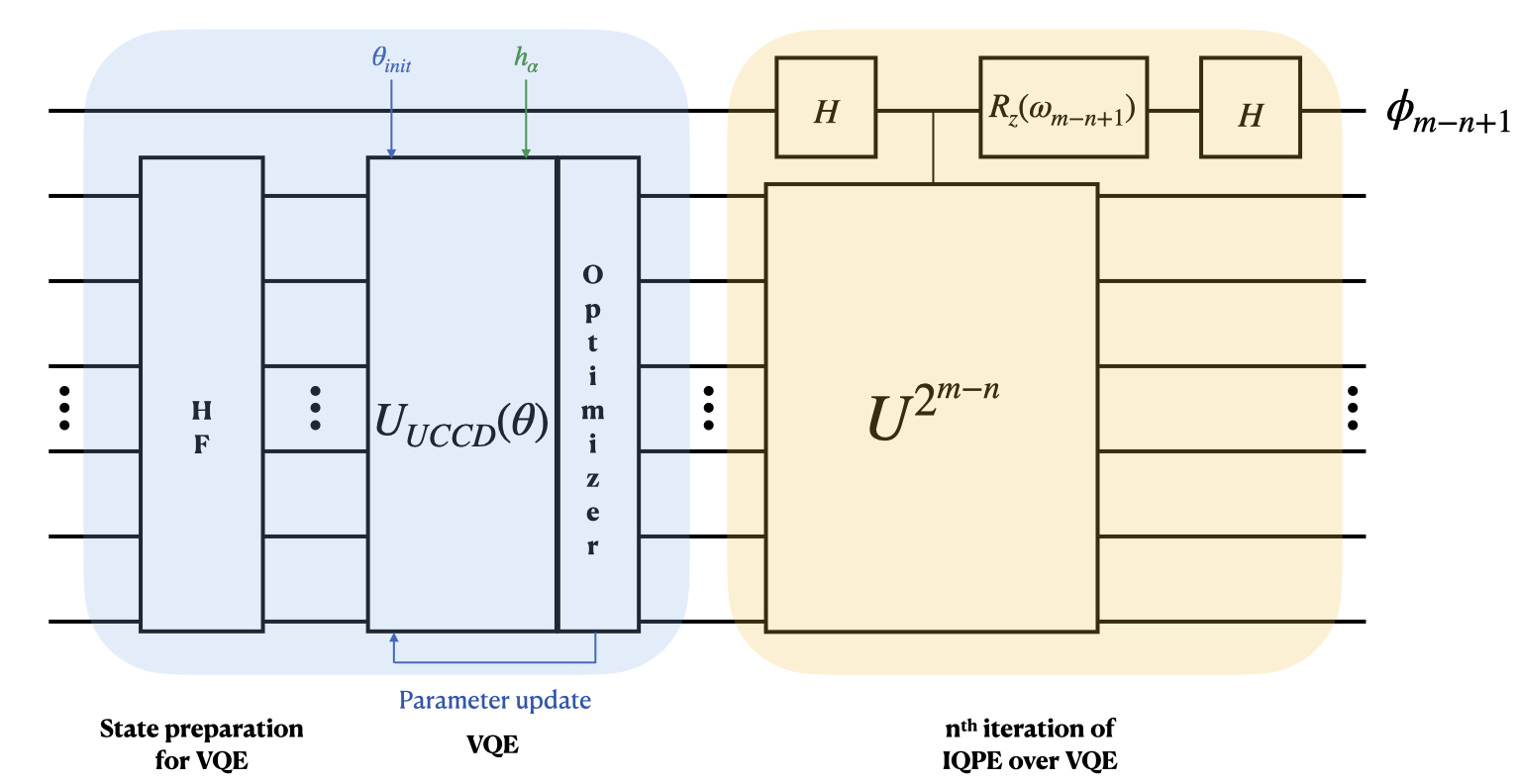} \\
            \textbf{(a)}&\textbf{(b)}\\
        \end{tabular}
\caption{An overview of the algorithm employed in this work. The choice of system is $H_4$ on a circle is shown in subfigure (a). Our scheme to obtain the ground state energy involves carrying out a variational quantum eigensolver (VQE) computation, followed by iterative quantum phase estimation (IQPE), as shown in subfigure (b). In the former, the unitary coupled cluster variational form with double excitations, denoted by $U_{UCCD}(\theta)$, is used. $\theta$ refers to the variational parameters, with the initial guess parameters given by $\theta_{init}$. $U_{UCCD}(\theta)$ acts on the HF initial state. $h_{\alpha}$ refers to the one- and two- electron integrals. The parameters are updated in each evaluation, and the multi-determinantal state thus prepared after optimization is fed into the IQPE algorithm. The figure shows a schematic of the algorithm for the $n^{th}$ iteration. }
\label{one}
\end{figure*}

\subsection{Methodology} 

We obtained the one- and two- electron integrals from PySCF~\cite{pyscf}, and carried out the VQE as well as the IQPE computations on Qiskit 0.26~\cite{qiskit}. We chose 1.738 $\AA$  for the radius of the $H_4$ circle as shown in Fig.~\cite{one}(a), the same as that from Ref.~\cite{h4}. We used the GAMESS~\cite{gamess} package (specifically, the determinant-based ALDET approach~\cite{aldet1,aldet2,aldet3}) for the relevant FCI coefficients to obtain success probabilities. \\ 

For our computations, we chose the contracted STO-3G basis~\cite{sto} that contains eight spinorbitals in total, COBYLA optimizer~\cite{cobyla}, Jordan-Wigner mapping scheme~\cite{map}, and the UCCD variational form~\cite{ucc}. It is worth adding at this point that UCCD with an untruncated double excitation space for $H_4$ on a circle in the STO-3G basis, leads to an 18-parameter VQE computation. We considered two options for $\theta_{init}$: zero initial guess, and the parameters built from many-body perturbation theory taken to second order in energy (MP2). We also worked with two choices for the backend simulator: the statevector and the QASM backend. We set a Trotter step of one for all our calculations. Also, we have employed the direct spinorbital to qubit mapping: that is each spinorbital is represented as one qubit, independent of the occupancy of the spinorbitals. \\

\subsection{Application to $H_4$ on a circle} 

The case of four $H$ atoms on a circle is a standard and challenging test bed to verify the efficacy of newly developed electronic structure algorithms. The system is characterized by the angle, $\beta$, as shown in Fig. \ref{one}(a). For values of $\beta$ away from 90\degree, for example, 80\degree, the system behaves as two non-interacting $H_2$ molecules, and is hence a strongly single reference case, with the correlated ground state wavefunction predominantly characterized by the HF state. This is clearly seen in the following FCI ground state generated via a  determinant-based CI approach: 

\begin{eqnarray}
\ket{\Psi_{FCI}^{80^{\degree}}}&=&0.642\ket{00110011}+0.350\ket{11001100} \nonumber \\ &-&0.317\ket{01010101}+0.251\ket{01101001} \nonumber \\&+&0.251\ket{10010110}-0.241\ket{10101010} \nonumber \\&+&0.223\ket{01011010}+0.223\ket{10100101} \nonumber \\&-&0.200\ket{10011001}-0.194\ket{0110 0110}.
\label{fci_80}
\end{eqnarray}

In the above equation (and the subsequent Eqs. (\ref{fci_89.8}), (\ref{vqe_80_500}) and (\ref{vqe_89p8_3000})), only those determinants are included which have absolute coefficients greater than 0.05.  Also, throughout this manuscript, each determinant has been written in its occupancy number representation, and the indices are arranged in the block spin format, where all the $\alpha$ spins are first counted, followed by the $\beta$ spins. From Eq. (\ref{fci_80}), we immediately see that the HF configuration dominates, followed by two doubly excited determinants. \\

Beyond 85\degree, the system starts showing the signature of molecular strong correlation, with one or more determinants becoming quasi-degenerate to the HF determinant. Very close to 90\degree, the HF determinant and a doubly excited determinant become
equally dominant, as shown in the FCI wavefunction below: 
\begin{eqnarray}
\ket{\Psi_{FCI}^{89.8^{\degree}}}&=&0.519\ket{00110011}-0.509\ket{01010101} \nonumber \\&+&0.345\ket{11001100}-0.340\ket{10101010} \nonumber \\&+&0.280\ket{01101001}+0.280\ket{10010110} \nonumber \\&+&
0.143\ket{10100101}+0.143\ket{01011010} \nonumber \\&-&0.137\ket{11000011}-0.137\ket{00111100}. 
\label{fci_89.8}
\end{eqnarray}

From the above equation, we see that at $89.8\degree$, the HF state and the doubly excited determinants are equally dominant in the description of the correlated exact ground state with $C_{HF}^{89.8\degree,GS}=0.519$ and $C_{01010101}^{89.8\degree,GS}=0.509$, and hence one needs to treat all the relevant dominant determinants in the same footing in order to have a balanced description of dynamical and static correlation effects. Furthermore, there are some excited states (denoted as $ES_n$) with same symmetry as the ground state (denoted as GS), where the coefficient of the HF determinant is sufficiently large in the corresponding FCI wavefunction, along with other doubly excited determinants. For example, for the lowest excited state $ES_1$, the coefficient of the HF determinant is -0.297 in the FCI wavefunction, whereas there exist two other excited states with the same symmetry (denoted as $ES_2$ and $ES_3$) where the HF determinant contributions are 0.543 and 0.431, respectively. Thus, starting from the HF state, the IQPE algorithm is probabilistically more likely to land up in the excited roots, as we will show in the subsequent section. This is quite unlike the scenario when $\beta=80\degree$. 
In our approach to follow, we would employ iterative optimization via the VQE-UCCD ans\"{a}tz to construct the reference determinant, and we would demonstrate that the overlap of the resultant wavefunction with the FCI wavefunction described in 
Eq. (\ref{fci_80}) and Eq. (\ref{fci_89.8}) can systematically be improved, which leads to increase the probability of success for the subsequent IQPE. Note that the ground state of $H_4$ system is of $A_g$ symmetry. It is clear that there is no single excitation operator belonging to the $A_g$ symmetry, and therefore, the inclusion of single excitation operators has no effect on the quality of the UCC wave function. \\

The multi-determinantal state thus prepared using the VQE algorithm is subsequently supplied as an input to the IQPE algorithm for phase estimation as shown in Fig. \ref{one}(b). One may note that with a larger number of VQE iterations, the initial reference state for IQPE gets a more substantial overlap with the exact target function. Thus, with more VQE iterations, the success probability of the IQPE algorithm to extract the correct phase is expected to increase. For a fixed number of IQPE iterations, the precision of an extracted phase remains unchanged. \\

\subsection{The potential energy surface and the effect of the number of IQPE iterations}

\begin{figure}[!t]
\includegraphics[height=50mm,width=70mm]{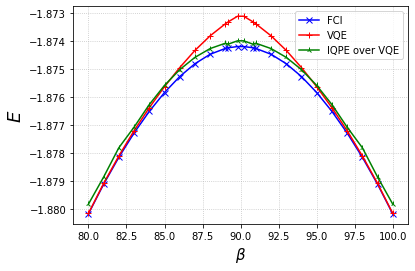}
\caption{Potential Energy Surface of $H_{4}$ on a circle, versus the angle $\beta$. The energies are given in units of Hartree ($E_h$). The blue curve gives the FCI values over the chosen range of angles, and the converged VQE results are provided in red. The results with 16 IQPE iterations over a state prepared by 1000 VQE iterations (via COBYLA optimizer and MP2 initial guess) are given in green color. The figure clearly shows that the non-parallelity error is very small in the IQPE over VQE method, whereas the VQE approach shows a marked deviation in the strong correlation regimes. }
\label{two}
\end{figure}

As a first step, we determine the PES of $H_4$ on a circle. In particular, we are interested in investigating the performance of our approach in both weak and strong correlation regimes. For this purpose, we plot, as shown in Fig.~\ref{two}, the ground state energies against the angle, $\beta$, which was defined in Fig.~\ref{one}(a). We vary $\beta$ from 80$\degree$ to 100$\degree$. Note that the plot is expected to be symmetric on both sides of $\beta = 90\degree$; however, due to inaccessibility of
the electronic configurations at $\beta=90\degree$, we stop at a value that is reasonably close to it~\cite{h4}. The plot also shows the FCI energies in the chosen range of angles, as it acts as the \textit{exact} result obtainable within a single particle basis, and hence a benchmark for comparison. We will elucidate on two aspects: precision (with respect to FCI) and non-parallelity error (NPE). It is evident from the results shown in the Fig.~\ref{two} that VQE with UCCD ans\"{a}tz performs very well and agrees to $\sim$ 0.1 milliHartree ($mE_h$) with respect to FCI, between $\beta=$ 80$\degree$ and $\beta=$ 85$\degree$, but starts deviating beyond that point. This behaviour shows that in geometries where strong correlation effects begin to become important, VQE no longer can make predictions of ground state energies with the kind of precision that it demonstrates for weakly correlated domains. In line with this observation, NPE provides a proper quantifier to assess the efficacy of a theory to treat the entire PES with same level of precision. NPE, across a PES, is defined as the difference between the maximum and minimum deviation of the results from a theory with respect to FCI~\cite{Sherrill1}. That is, it quantifies the spread of the (non-)parallelity error across the PES. We find from our data that the NPE is more than 1 $mE_h$ for VQE when UCCD ans\"{a}tz is employed, which is not within acceptable bounds, given that the chemical accuracy itself is $\sim$ 1 $mE_h$. To check if IQPE over VQE circumvents the large NPE obtained from VQE, we employed the IQPE algorithm with 16 iterations, with the initial reference state prepared using VQE in the UCCD ans\"{a}tz with 1000 iterations, and using MP2 initial guess parameters. Our calculation show that the NPE was only about 0.2 $mE_h$ for IQPE when started with the reference function generated by VQE with UCCD ans\"{a}tz. This is not surprising as a 16-bit IQPE is supposed to provide energy, which is precise up to $\sim 10^{-4} E_h$, and thus NPE is, in principle, controlled only by the number of iterations in IQPE. However, with VQE-UCCD, the quality of the initial guess wave function greatly affects the probability to obtain the correct ground state in the subsequent IQPE stage. We will elucidate this point in the next subsection. \\ 

\begin{figure}
\centering
    \setlength{\tabcolsep}{1mm}
        \begin{tabular}{c}
            \includegraphics[height=50mm,width=70mm]{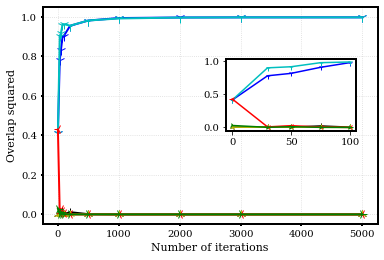} \\
            \textbf{(a)}\\
            \includegraphics[height=50mm,width=70mm]{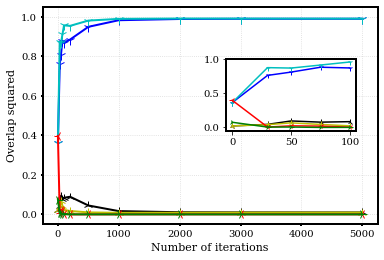} \\
            \textbf{(b)}\\
            \includegraphics[height=60mm,width=70mm]{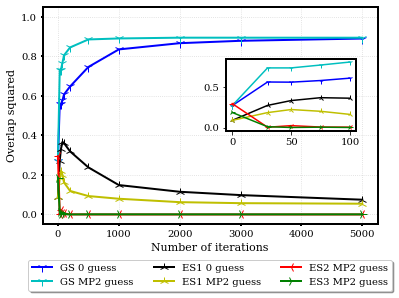} \\
            \textbf{(c)}\\
        \end{tabular}
    \caption{Plots illustrating our results for three chosen angles, $\beta=$ 80\degree (subfigure (a)), $\beta=$ 85\degree (subfigure (b)), and $\beta=$ 89.8\degree (subfigure (c)), reflecting the increasing importance of strong correlation effects. Each of the sub-figures contain the overlaps of both the ground and excited states with the FCI state, defined by $\langle \Psi \arrowvert \Psi_{FCI} \rangle$, plotted against the number of VQE iterations for a given angle. The blue and black curves denote the ground (GS) and an excited state (ES1) overlaps respectively, with zero initial guess for the VQE parameters. The cyan and yellow curves present the same quantities, but for an MP2 initial guess for the VQE parameters. Additionally, we also provide the overlaps for two other excited states (ES2 and ES3), starting with MP2 guess, and they are shown as red and green curves, respectively.  } 
\label{four}
\end{figure}

We should also note that IQPE over a HF reference determinant is in most occasions unable to extract the correct phase that corresponds to the ground state eigenfunction for all angles beyond 80$\degree$ (\textit{vide infra}). Since the HF determinant has significant contribution to the excited state wavefunctions as compared to the ground state wavefunction, as described in the previous section and also shown in Fig. \ref{four}, starting from the HF reference determinant in most occasions leads to incorrectly landing on an excited state eigenfunction. It is also worth adding that such a quantitatively accurate PES obtained from IQPE over VQE with high probability of success can later be extended to evaluating properties such as vibrational frequencies, dissociation energies, etc. This attests the strength of IQPE over VQE, provided the initial state from VQE is chosen well. \\

\begin{figure}[t]
\includegraphics[height=50mm,width=70mm]{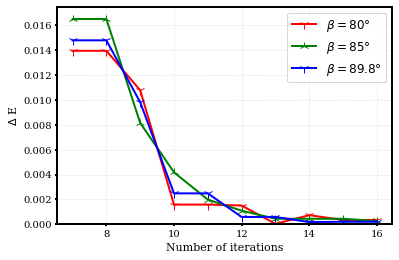}
\caption{Difference between FCI and the IQPE energies (in $E_h$) versus the number of IQPE iterations, for three geometries: (a) $\beta = 80 \degree$, (b) $\beta = 85 \degree$, and (c) $\beta = 89.8 \degree$. The initial reference function was optimised via VQE algorithm with 1000 COBYLA iterations, starting from MP2 guess amplitudes.}
\label{three}
\end{figure}

To understand the number of IQPE iterations at which the calculated energy converges for $H_4$ on a circle, we have plotted in Fig.~\ref{three} the change in the difference between the FCI energy and observed energy with number of IQPE iterations, for $\beta = 80 \degree,\ 85 \degree,\ \mathrm{and} \ 89.8 \degree$. We have used MP2 initial guess for VQE parameters, and set maximum number of VQE iterations to 1000 for the state preparation. It is worth adding at this point that we have not gone beyond 16 IQPE iterations in view of the increasingly steep computational requirements. The figure shows that one needs to go at least upto 13 IQPE iterations to reach sub-$mE_h$ precision, independent of whether we are in weak or strong correlation regime. This is not surprising as a 13-bit IQPE calculation leads to precision of $\pm (2\pi\times 2^{-13}) E_h \sim 10^{-4} E_h$; it is simply the probability of correctly extracting the phase to a pre-defined precision, while noting that the value on which we land can greatly be enhanced in a controlled manner via tuning various VQE parameters. The importance of an appropriate state preparation is far more obvious for strong correlation regimes than single reference regimes, which we would demonstrate later.\\

\subsection{The effect of the number of VQE iterations: state preparation for IQPE}

The success of the IQPE algorithm depends on the \textit{closeness} of the initial reference state to the exact eigenstate of the many-body Hamiltonian. Although the final precision of the estimated energy by the IQPE algorithm 
is independent of the initial reference state, an appropriately prepared reference state largely enhances the probability of
extracting the correct phase corresponding to the target state. In our approach, starting from the HF determinant, we have constructed a multi-determinantal reference state by including dynamical correlation through iterative optimization via the VQE algorithm. 
Understandably, a sufficiently large number of VQE iterations would ensure that the resulting multi-determinantal state has substantial overlap with the exact ground eigenstate. Furthermore, it is observed that if the initial state is poorly prepared (e.g.
a HF state, or a multi-determinantal state generated with a very low number of VQE iterations with zero or random 
initial guess amplitudes), there is a non-negligible probability
that IQPE extracts a phase which corresponds to an excited state. We find that this is particularly the situation for molecules in strongly correlated regime since the UCCD ans\"{a}tz often fails to account for the short range static correlation. To this end, we have presented the squared overlap of the prepared state, with the FCI ground (denoted as GS)
as well as with lowest excited (denoted by ES$_n$) states, as functions of the
number of VQE iterations for three different angles arranged in increasing order of strong correlation, as shown in Fig.~\ref{four}. In the weak correlation regime, for which we have chosen 80$\degree$ as the representative case, the HF determinant has a squared overlap with the FCI ground state of nearly $|\langle \Psi_{HF} | \Psi_{FCI}^{80\degree,GS}\rangle|^{2}=0.41$, while its squared overlap with the lowest excited state, $|\langle \Psi_{HF}|\Psi_{FCI}^{80\degree,ES_1}\rangle|^{2}=0.004$. We note that we have introduced an additional index on the superscript 
wherever necessary, to distinguish between GS and ES FCI states. With only a few VQE iterations, the overlap between the reference 
state and the target FCI ground state can significantly be improved: In fact, with only about 500 iterations and a single 
VQE calculation starting from MP2 amplitudes, the correlated state takes the form 
\begin{eqnarray}
\ket{\Psi_{VQE(500)}^{80^{\degree}}}&=&0.639\ket{00110011}+0.356\ket{11001100} \nonumber \\ &-&0.322\ket{01010101}+0.211\ket{01101001} \nonumber \\&+&0.211\ket{10010110}-0.244\ket{10101010} \nonumber \\&+&0.255\ket{01011010}+0.254\ket{10100101} \nonumber \\&-&0.188\ket{10011001}-0.184\ket{0110 0110}\nonumber \\&-&0.073\ket{00111100}-0.074\ket{11000011},\nonumber \\
\label{vqe_80_500}
\end{eqnarray}
having a squared overlap of nearly one with the FCI ground state (shown 
in Eq. (\ref{fci_80})), that is,  $|\langle \Psi_{VQE(500)}^{80\degree} |\Psi_{FCI}^{80\degree,GS}\rangle|^{2} \rightarrow 1.0$, while $|\langle \Psi_{VQE(500)}^{80\degree} |\psi_{FCI}^{80\degree,ES_n}\rangle|^{2} \rightarrow 0$. This ensures 
guaranteed estimation of the correct phase corresponding to the ground state. On the other hand, in the strongly correlated regime (89.8$\degree$), the HF determinant has significant squared overlap with both the ground and a couple of excited states of the same symmetry, and hence the phase estimation with HF state as the initial reference often incorrectly leads to one of the excited states. \\

This issue can greatly be alleviated by optimising the reference determinant via the inclusion of correlation effects through VQE; the overlap between the VQE state and the FCI ground state steadily increases with the number of VQE iterations and it saturates around 1000 VQE iterations when we begin with either the MP2 or the zero initial guess amplitudes. Further optimization does not lead to any improvement of the VQE energy and the prepared state. As an example, 
\begin{eqnarray}
\ket{\Psi_{VQE(3000)}^{89.8^{\degree}}}&=&0.565\ket{00110011}-0.407\ket{01010101} \nonumber \\&+&0.378\ket{11001100}-0.270\ket{10101010} \nonumber \\&+&0.262\ket{01101001}+ 0.262\ket{10010110}\nonumber \\&+&0.193\ket{10100101}+0.193\ket{01011010} \nonumber 
\\&-&0.071\ket{11000011}-0.071\ket{00111100}\nonumber
\\&-&0.124\ket{10011001}-0.120\ket{01100110}.\nonumber\\
\label{vqe_89p8_3000}
\end{eqnarray}
This resulting multi-determinantal state, generated with 3000 
iterations by a single VQE calculation starting from the MP2 
amplitudes, has about 95\% overlap with the 
FCI ground state: $|\langle \Psi_{VQE(3000)}^{89.8\degree} |\Psi_{FCI}^{89.8\degree,GS}\rangle|^{2} =0.9$. On the other hand,
its overlap with the lowest excited state 
reduces significantly to $|\langle \Psi_{VQE(3000)}^{89.8\degree}|\psi_{FCI}^{89.8\degree,ES_1}\rangle|^{2}=0.057$. Starting from zero guess amplitudes, we find that $|\langle \Psi_{VQE(3000)}^{89.8\degree} |\Psi_{FCI}^{89.8\degree,GS}\rangle|^{2} =0.88$, and $|\langle \Psi_{VQE(3000)}^{89.8\degree}|\psi_{FCI}^{89.8\degree,ES_1}\rangle|^{2}=0.096$. One may also note that the overlap of the VQE generated
state with $ES_2$ and $ES_3$ sharply decays to nearly zero with 
very few VQE iterations, as shown in Fig. \ref{four}(c). We argue
that the construction of the multi-determinantal reference state 
via VQE iterations leads to significantly enhanced overlap with the target ground state, ensuring a quicker and secured evolution via IQPE. \\

\begin{figure*}[t]
\centering
\includegraphics[width=\textwidth]{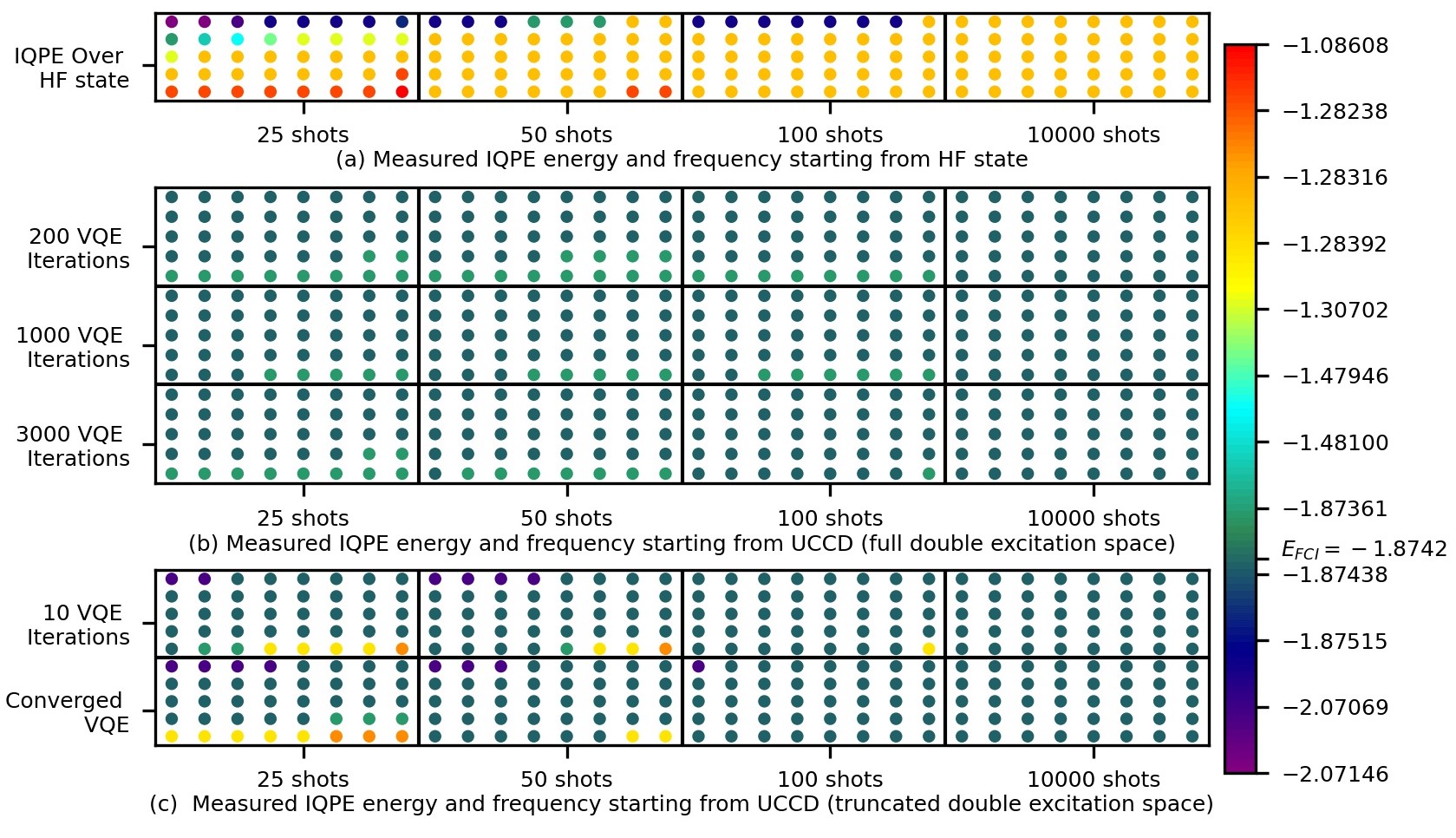}
\caption{Figure illustrating the dependence of the quality of our results on the VQE iterations (a measure of the goodness of the initial state for IQPE procedure) as well as on the number of shots for 89.8\degree. The energies are appropriately color coded, as shown on the right panel. The exact FCI energy is also marked appropriately in the panel. Each circle denotes one repetition of the experiment for a given number of shots and a given initial state preparation. The number of same colored circles within a block, therefore, reflects the frequency of occurrence of a given value of energy. Note the appearance of two different colours (in subfigure (b) with 25, 50 and 100 shots, corresponding to two different closely spaced energy values obtained by IQPE when the initial reference state is constructed via full double excitation inclusive VQE. For the extreme case of 10000 shots, IQPE predicts a single value for the energy, irrespective of the truncation scheme in UCCD ans\"{a}tz for the VQE state preparation. }
\label{five}
\end{figure*}

We briefly comment on the error due to Trotter step, and found it 
to be well within 0.1 $mE_h$. We also checked the error from the
choice of mapping (Jordan-Wigner, Parity, and Bravyi-Kitaev 
schemes), and found that too to be well within 0.1 $mE_h$. This 
is not too surprising, as the transformed qubit Hamiltonians from
each of these mappings are expected to be isospectral to the
electronic Hamiltonian~\cite{iso}. \\ 

\subsection{Effect of IQPE state preparation on noise due to sampling}

The results discussed until now employ the SV backend. We now study the influence of the number of individual bit sampling (also referred to as shots) for IQPE in our results, by using IBM's QASM backend. It is well known that in IQPE, the probability of correctly extracting any individual bit is, in practice, always less than one due to inexact phase expansion. One way to alleviate this is to repeat the sampling of each bit multiple times, and obtain the correct phase via the classical error correction technique. If the probability of extracting a given bit is greater than 0.5, repeated sampling would reduce the probability of the measurement error exponentially~\cite{Chernoff}. In the following paragraphs, we investigate the effect of sampling noise in IQPE, with the state preparation using HF state and the VQE algorithm. The VQE stage uses the UCCD ans\"{a}tz, which is set via two different truncation schemes. In the first scheme, we use an untruncated double excitation space, whereas for the second, we truncate the double excitation space, and term the latter as minimally parametrized UCCD ans\"{a}tz. For all the computations discussed below, we choose the representative case of 89.8\degree with 14-bit IQPE based on our observations from Fig.~\ref{three}, and vary the number of VQE iterations and the number of shots together. We use MP2 initial guess for the VQE calculations. Each such computation is repeated 40 times to build reasonable statistics. 

\subsubsection{State preparation via the HF state} 

While starting with the HF state as the initial state for IQPE (the traditionally followed IQPE approach) as shown in Fig.~\ref{five}(a), the IQPE algorithm almost always estimates the wrong eigenphase. Starting from the HF determinant and with as low as 25 shots, the measured eigenphases show severe spread ($\Delta$, which we define as the difference between the highest and the lowest observed values) of 1.60 $E_h$ when the experiment is repeated 40 times. As we increase the number of shots, the spread of the measured eigenvalues reduces ($\Delta=$0.985 $E_h$ and 0.788 $E_h$ with 50 and 100 QASM shots, respectively), and eventually at a sufficiently large number of shots (10000 shots), leads to a single measured value of the eigenstate, which corresponds to one of the \textit{excited roots}. This result is consistent with our findings from calculations using the SV backend. 

\subsubsection{State preparation via the UCCD ans\"{a}tz with untruncated double excitation space}

In this section, we briefly discuss our findings when IQPE is repeated with the initial reference state generated via VQE-UCCD with complete double excitation manifold. The reference state preparation thus involves a 18-parameter VQE computation. 
The results shown in Fig.~\ref{five}(b) reflect that an appropriately prepared multi-determinantal state with \textit{complete} inclusion of two-body excitation operators in the UCCD ans\"{a}tz can greatly reduce the requirement of repeating the experiment with large number of QASM shots (in the IQPE stage of the algorithm). In contrast to the trend observed when the initial state for IQPE is the HF determinant, the current choice of initial state displays a much lesser spread in energies. Even with a low number of QASM shots, but starting with a reasonable state (prepared with 200 VQE iterations), we observe a spread of only a few $mE_h$ ($\Delta=$0.00076 $E_h$) around the exact ground state energy (Fig.~\ref{five}(b)). This is already comparable to the desired tolerance limit, given the 14-bit IQPE achieves the target precision of $\pm (2^{-14} \times 2\pi) E_h \sim 10^{-4} E_h$. With further increase in the number of shots for IQPE, the estimated eigenphase shows high bias towards the modal value, which corresponds to the \textit{correct} ground state energy that one observes using the SV backend. Increase in the number of VQE iterations shows little effect on the spread of the observed energy values, although it enhances the bias towards the modal value, particularly when the number of QASM samplings per individual bit is increased. We note that one may obtain the correct eigenphase with much lower number of VQE iterations and the number of individual bit samplings in IQPE (QASM shots) when the VQE calculation is performed starting from the MP2 guess amplitudes, than starting from zero. Nonetheless, even starting from zero guess amplitudes, upon 40 repetitions of the experiment, one obtains a sub-$mE_h$ spread with very low numbers of QASM shots and VQE iterations. Thus, we numerically demonstrate that an appropriately prepared state leads to significant increase in the probability of correctly obtaining the desired phase to a given precision.  \\

\subsubsection{The reference state preparation with minimally parametrized UCCD ans\"{a}tz}

While the UCCD ans\"{a}tz employed above for the reference state preparation is quite  promising, it may often be plagued by  incomplete inclusion of static correlation in the regions of molecular strong correlation. Furthermore, UCC in the entire double excitation space involve variational optimization of a large number of parameters, which results in higher cost. Since the static correlation often plays a dominant role over dynamical correlation for generating 
the wave function having larger overlap with the FCI wavefunction, one may selectively include only those cluster operators in the unitary ans\"{a}tz which are labelled by the \textit{chemically active} orbitals. This reduces the number of variational parameters drastically while includes the static correlation effects to a high extent. We thus propose:
\begin{equation}
    U_{min} = \exp{(T_{act}(\theta)-T_{act}^\dagger(\theta))}
\end{equation}

where $T_{act}$ is the subset of two-body excitation operators, which are labelled only by the active orbitals. Below, we shall demonstrate that the minimally parametrized ans\"{a}tz proposed above is quite accurate for the purpose of reference state preparation
with efficient inclusion of the static correlation effect for future study of QPE-based approaches. 

Now, we systematically study the effect of noise due to sampling when the initial reference state is prepared via VQE but within a truncated excitation space (minimally parametrized ans\"{a}tz). For this purpose, we again chose to work in the strong correlation regime ($\beta=$89.8\degree). We note from Eq. (\ref{fci_89.8}) that the exact ground state wavefunction has large contribution from two dominant determinants: the HF determinant, and the doubly excited determinant, where two electrons from the highest occupied molecular orbital (one from the alpha spinorbital and one from beta spinorbital) are excited to the lowest unoccupied molecular spinorbitals. Truncation of the excitation space thus leads to the reduction of the number of VQE parameters to only one. In such a case, the UCCD ans\"{a}tz reduces to $U_{min}=\exp(T_{15}^{26}-(T_{15}^{26})^\dagger)$, with $T_{15}^{26}=\theta_{15}^{26}\{a_2^\dagger a_6^\dagger a_5 a_1\}$, where $a_2^\dagger$ $(a_1)$ and $a_6^\dagger$ $(a_5)$ are the creation (annihilation) operators with up and down spinorbitals in our block spin arrangement. Note that the iterative optimization of the single VQE parameter generates the multi-determinantal state which is a linear superposition of $\ket{00110011}$ and $\ket{01010101}$. The obtained results from Fig. \ref{five}(c) show that IQPE phase estimation starting from the multi-determinantal state prepared within the truncated active excitation space shows similar statistics upon repeated experiments as we previously obtained when started from a multi-determinantal reference state generated via the complete space VQE optimization. 

In Fig. \ref{five}(c) bottom panel, we show that with the converged VQE parameters (where convergence was reached at 24 iterations; note that this is in contrast to full excitation space case, where convergence is reached at well over 3000 VQE iterations) and as few as 25 QASM shots for IQPE bit measurement, the final energy shows significantly large spread ($\Delta=0.98 E_h$) compared to when the initial reference state was prepared with untruncated doubles excitation UCCD. However, the modal value (with frequency 25 times out of 40 repetitions) is same as one observes with SV or using a full excitation space UCCD. As expected, with more number of shots, the spread in the observed energy decreases, with a sharper gradient than one observes with untruncated UCCD states. Moreover, the modal value shows stronger bias towards the exact one observed with SV backend (with frequency 35, 39 and 40 times out of 40 repetitions with 50, 100 and 10000 QASM shots respectively). Furthermore, as shown in Fig. \ref{five}(c) top panel, a similar statistics is observed when the optimization of the reference state over the single parameter is restricted to only 10 iterations. Thus, due to the computational gain and given the constraint in resources in the current era, we recommend for a state preparation with truncated excitation space UCCD ans\"{a}tz over a complete space UCCD for a subsequent evolution by IQPE. 

\section{Conclusion}\label{conclusion}

In this work, we presented a hybrid IQPE over VQE algorithm for the digital quantum simulation of molecular strong correlation effects in $H_4$ on a circle. While the VQE algorithm using the UCCD ans\"{a}tz does not perform satisfactorily around geometries where molecular strong correlation becomes important, as exemplified by the non-parallelity of the PES as compared to its FCI counterpart, the hybrid IQPE over VQE algorithm reproduces molecular PES with very good precision regardless of the electronic complexity. In particular, we find a whole order of magnitude improvement at strongly correlated regimes. 

With a reasonably high number of initial VQE iterations, one may prepare a multi-determinantal reference state that has substantial overlap with the exact ground eigenstate of the many-body Hamiltonian. This results in a much faster and safer evolution via the subsequent IQPE algorithm. We find that starting from a well prepared multi-determinantal state, the IQPE algorithm takes more or less the same number of iterations for an acceptable precision in energy, across different geometries. While the final precision of the estimated ground state energy is determined by the number of IQPE iterations, the success probability of landing in the correct eigenstate is controlled by the number of VQE iterations. We also checked that one may further tune the initial guess parameters at the VQE step to further increase the success probability. \\

We also studied the effect of sampling noise by using the QASM backend for a geometry that displays strong correlations in $H_4$ on a circle. We reported results on three distinct cases for the choice of initial state for IQPE, namely HF state, full excitation space UCC, and minimally parametrized UCC. The HF result lands on an excited root, and was consistent with the SV results. Although the trends with increasing number of shots are dissimilar for the cases that use full excitation space UCC and minimally parametrized UCC, they produce the same results when VQE is converged and IQPE is run with a large number of shots. Our findings point that the bias of the modal value can systematically be tuned with goodness of the prepared state, where relevant excitations can be chosen with  physically/chemically motivated arguments, and the estimated modal value can show strong bias towards the exact ground state energy upon repetition of the experiment multiple times. Our detailed pilot study with the hybrid IQPE over VQE algorithm can serve as a stepping stone to future explorations of strong correlation effects using quantum computers. 

\section*{Acknowledgements}

We are very thankful to Prof. K. Sugisaki for stimulating discussions on the conceptual underpinnings of the IQPE algorithm. We thank Prof. B. P. Das and Prof. D. Mukherjee for useful discussions on quantum computing in the initial stages of this work. Most of the calculations were performed on the computational facilities at IIT Bombay. The rest of the calculations were carried out on National Supercomputing Mission's (NSM) computing resource, `PARAM Siddhi-AI', at C-DAC Pune, which is implemented by C-DAC and supported by the Ministry of Electronics and Information Technology (MeitY) and Department of Science and Technology (DST), Government of India. RM thanks Science and Engineering Research Board DST, Government of India and IRCC, IIT Bombay for funding.\\

\end{document}